\documentclass[]{interact}

\usepackage{epstopdf}
\usepackage[doublespacing]{setspace}
\usepackage{multirow}
\usepackage{multicol}
\usepackage{hyperref}
\usepackage{xcolor}
\usepackage{longtable}
\usepackage{graphicx}
\usepackage{caption}
\usepackage{subcaption}

\usepackage[natbibapa,nodoi]{apacite}
\theoremstyle{plain}

\theoremstyle{definition}

\theoremstyle{remark}

\begin{document}
\setstretch{1}
\section*{Multilevel mixtures of latent trait analyzers for clustering multi-layer bipartite networks}

Dalila Failli \textsuperscript{a}\thanks{CONTACT Dalila Failli. Email: dalila.failli@unifi.it}, Bruno Arpino \textsuperscript{b}\thanks{CONTACT Bruno Arpino. Email: bruno.arpino@unipd.it}, Maria Francesca Marino \textsuperscript{a}\thanks{CONTACT Maria Francesca Marino. Email: mariafrancesca.marino@unifi.it}\\
\affil{\textsuperscript{a} Dipartimento di Statistica, Informatica, Applicazioni, Università degli Studi di Firenze, Firenze, Italy; \textsuperscript{b}Dipartimento di Scienze Statistiche, Università di Padova, Padova, Italy}
\vspace*{5mm}

\begin{abstract}
Within network data analysis, bipartite networks represent a particular type of network where relationships occur between two disjoint sets of nodes, formally called sending and receiving nodes. In this context, sending nodes may be organized into layers on the basis of some defined characteristics, resulting in a special case of multi-layer bipartite network, where each layer includes a specific set of sending nodes. To perform a clustering of sending nodes in multi-layer bipartite network, we extend the Mixture of Latent Trait Analyzers (MLTA), also taking into account the influence of concomitant variables on clustering formation and the multi-layer structure of the data. To this aim, a multilevel approach offers a useful methodological tool to properly account for the hierarchical structure of the data and for the unobserved sources of heterogeneity at multiple levels. A simulation study is conducted to test the performance of the proposal in terms of parameters’ and clustering recovery. Furthermore, the model is applied to the European Social Survey data (ESS) to i) perform a clustering of individuals (sending nodes) based on their digital skills (receiving nodes); ii) understand how socio-economic and demographic characteristics influence the individual digitalization level; iii) account for the multilevel structure of the data; iv) obtain a clustering of countries in terms of the base-line attitude to digital technologies of their residents. 
\end{abstract}

\begin{keywords}
Model-based clustering, Network data, Concomitant variables, Multilevel analysis, EM algorithm, Variational inference
\end{keywords}

\section{Introduction}
\label{sec:intro}
In many research fields, data can be naturally depicted as structures of interconnected nodes, known as networks. A particular type of network is the bipartite network, representing connections between two distinct sets of nodes, called sending and receiving nodes, respectively.\\This type of network is particularly useful for representing various phenomena, including connections between company directors and boards of directors (see e.g. Davis et al., 2003), between countries and exported products (see e.g. Hidalgo \& Hausmann, 2009), as well as between genes and protein products (see e.g. Pavlopoulos et al., 2018), and between patients and measures adopted to avoid infection during a pandemic (see e.g.Failli et al., 2022).\\In some contexts, sending nodes are naturally organized into layers on the basis of some predefined characteristics, resulting in a special case of multi-layer bipartite network. In such a network, several layers exist, each comprising a specific type of sending nodes, which are in turn connected to receiving nodes. For instance, this is the case of multi-layer ecological networks, i.e., networks involving different types of animal species, which can be used to predict the indirect effects of species loss on multi-species assemblages (see e.g. Astegiano et al., 2017). Another area of application concerns the biological field, where multi-layer bipartite networks are employed for drug-target interaction prediction (see e.g. Koptelov et al., 2021). Multi-layer network can also be used in the economic research field. An example is given by Alves et al. (2019), who consider sellers and buyers as nodes of a bipartite network, and industries as layers, with the aim of identifying countries and sectors that contribute most to the transaction process.\\In this paper, we aim at performing a clustering of sending nodes in multi-layer bipartite networks by extending the Mixture of Latent Trait Analyzers (MLTA; Gollini \& Murphy, 2014; Gollini, 2020). This model allows to identify latent groups of nodes sharing unobserved characteristics, as in the latent class framework, as well as to capture the residual dependence of network connections, as in the latent trait framework. The combination of these features makes it possible to overcome the problem related to the local independence assumption upon which the latent class model is based (Bartholomew et al., 2011) and which may lead to the identification of an excessive number of groups, thus making the interpretation of results difficult. However, the original specification of this model does not account for the effect that covariates may have on clustering formation. To this aim, we follow the approach of Failli et al. (2022), who explicitly model the effect that concomitant variables may have on the clustering structure via a multinomial logit specification for the prior probabilities of the latent classes. Furthermore, we exploit a multilevel approach to properly account for the hierarchical structure of the data and for the unobserved sources of heterogeneity at multiple levels. To this aim, a {layer}-specific random effect is considered. Maximum likelihood estimates of model parameters can be obtained by maximizing the log-likelihood function via an EM algorithm (Dempster et al., 1977). Furthermore, the multidimensional integral to be solved in order to derive the log-likelihood function is approximated via a variational approach (Jaakkola \& Jordan, 1997).\\A simulation study is conducted to test the performance of the proposal in terms of parameters’ and clustering recovery. As far as the real-data application is concerned, the model is applied to the European Social Survey data round 10 (2020) to perform a clustering of individuals (sending nodes) based on their digital skills (receiving nodes), also taking into account the multilevel structure of the data (i.e., individuals nested into countries) and the influence of nodal attributes on the individual level of digitalization. The use of a non-parametric maximum likelihood (NPML; Laird, 1978) approach on the random effect also allows countries to be classified in terms of the baseline attitude to digital technologies of their residents.\\The paper is organized as follows. In Section 2we describe the ESS data. In Section 3, we introduce the assumptions underlying the original specification of the MLTA model. In Section 4, we describe the proposed extension of the MLTA model, also detailing parameters estimation and model selection. Section 5 shows the results of a simulation study conducted in order to verify the efficacy of the proposed approach in terms of parameters’ recovery and clustering. Section 6 presents the application of the proposed model to the European Social Survey data set. Section 7 contains concluding remarks and points to further extensions of the approach.

\section{The European Social Survey data}
\label{sec:application}
In this work, we focus on the European Social Survey (ESS) data from round 10, an academic transnational survey whose main objective is to analyse the evolution of attitudes in Europe. It covers 21 countries and includes questions on social and political behaviour, health, and family life, as well as a module based on digital social contacts in work and family life, upon which our analysis is developed. The countries considered in the analysis are:
\begin{itemize}
    \item Belgium
    \item Bulgaria
    \item Switzerland
    \item Czech Republic
    \item Estonia
    \item Finland
    \item France 
    \item United Kingdom
    \item Croatia
    \item Hungary
    \item Ireland
    \item Iceland
    \item Italy
    \item Lithuania
    \item Montenegro
    \item North Macedonia
    \item Netherlands
    \item Norway
    \item Portugal
    \item Slovenia
    \item Slovakia
\end{itemize}
These include all countries for which there are no missing data.\\
Motivated by the analysis of these data, we build a bipartite network related to the presence of various digital skills (receiving nodes) in individuals (sending nodes) aged 50+ from different European countries. Our goal is to identify groups of older adults with similar digital skills, also taking into account the multilevel structure of the data (i.e., individuals nested within countries) and the influence of socio-economic and demographic characteristics on the individual level of digitalization. As it has already been pointed out, the model specification also allows countries to be classified in terms of the baseline attitude to digital technologies of their residents.\\Abrendoth et al. (2023) show that the majority of respondents use the Internet every day, although 25\% of individuals never use it. In addition, a small percentage of respondents are completely familiar with preference settings, advanced search, and PDFs. With regards to video calls and text messages, most of respondents find these skills not applicable, as they do not have children aged 12 or over, nor do they have parents, or they do not communicate via video calls or messages for work. Furthermore, they show that digital contacts at work are not yet widely spread and that countries are generally characterised by more frequent communication with children than with parents.\\To build the multi-layer bipartite network, we assume that a relation between pair of nodes does exist if an individual often uses or is somewhat familiar with a certain digital skill. These indicate whether an individual possesses a particular digital competence, such as: 
\begin{itemize}
    \item Internet use;
    \item Preference settings use;
    \item Advanced search use;
    \item PDFs use;
    \item Video calls with children aged 12 or over, parents, manager, or colleagues;
    \item Messages (text, email or messaging apps) with children aged 12 or over, parents, manager, or colleagues;
    \item Sharing of online posts about politics in the last 12 months.
\end{itemize}
The distribution of such variables is shown in Table 1. From this table, we can see that 67\% of the respondents use the Internet, while less than 50\% of them use preference settings, advanced search and PDFs. The percentage of those who make video calls is 35\%, while about 50\% send messages. Finally, only 11\% of respondents share online political posts. It is worth noting that, although Internet use may be a pre-requisite for some other skills, it is not a pre-requisite for other digital competences (which only pre-suppose the use of a computer), and there are no filters in the questionnaire, meaning that some respondents may not include Internet use in the other competences.
\begin{table}[ht!]
\small
    \centering
\caption{Distribution of digital skills variables.}
\label{dig}
\begin{tabular}{l r}
\noalign{\smallskip}\hline\noalign{\smallskip}
    Variable  & Frequency\\
\noalign{\smallskip}\hline\noalign{\smallskip}
 Internet & 0.67  \\
\noalign{\smallskip}\hline\noalign{\smallskip}
 Preference settings & 0.47\\ 
\noalign{\smallskip}\hline\noalign{\smallskip}
 Advanced search & 0.49 \\
\noalign{\smallskip}\hline\noalign{\smallskip}
 PDFs & 0.47\\
\noalign{\smallskip}\hline\noalign{\smallskip}
 Video calls & 0.35\\
\noalign{\smallskip}\hline\noalign{\smallskip}
 Messages & 0.51\\
\noalign{\smallskip}\hline\noalign{\smallskip}
 Political posts & 0.11\\
\noalign{\smallskip}\hline\noalign{\smallskip}
 \end{tabular}
\end{table}
Considering only individuals with no missing records, we obtain a multi-layer bipartite network made of $13174$ sending nodes (individuals), $7$ receiving nodes (digital skills), and $21$ layers (countries).\\The data set also contains information on socio-demographic attributes, such as age, education, and health, as well as variables measuring the existence of intergenerational ties, such as the presence of children in the household. A detailed description of the variables is included in the Supplementary Materials. These variables are transformed into a suitable number of dummy variables or into variables with a smaller number of categories, as listed in Table 2, in order to use them as concomitant variables for the application of the proposed approach to the ESS data.
\begin{table}[ht!]
\small
    \centering
\caption{Distribution of demographic and socio-economic variables.}
\label{descrittive2}
\begin{tabular}{l l r}
\noalign{\smallskip}\hline\noalign{\smallskip}
    Variable     &  Category & Frequency\\
\noalign{\smallskip}\hline\noalign{\smallskip}
 Age & $50\vdash65$  & 0.49  \\
 & $65\vdash75$ &  0.31   \\
  & $75+$ & 0.20   \\ 
\noalign{\smallskip}\hline\noalign{\smallskip}
 Health & Bad & 0.12\\ 
&	Fair & 0.35\\ 
&	Good &  0.53\\
\noalign{\smallskip}\hline\noalign{\smallskip}
 Hampered in daily activities &Yes & 0.37 \\
 & No & 0.63\\
\noalign{\smallskip}\hline\noalign{\smallskip}
 Born in country & Yes & 0.92\\
 & No & 0.08 \\
\noalign{\smallskip}\hline\noalign{\smallskip}
 Gender & Male & 0.46\\
    & Female & 0.54\\
\noalign{\smallskip}\hline\noalign{\smallskip}
 Marital status & With partner & 0.58\\
&	Without partner & 0.42\\
\noalign{\smallskip}\hline\noalign{\smallskip}
 Education & Low & 0.27\\
 & Medium & 0.39\\
 & High & 0.35\\
\noalign{\smallskip}\hline\noalign{\smallskip}
 Partner education & Low & 0.15\\
& Medium & 0.25\\
& High & 0.21\\
& Not applicable & 0.39\\
\noalign{\smallskip}\hline\noalign{\smallskip}
 Household's total net income &  Low & 0.37\\
 &  Medium & 0.33\\
 & 	High & 0.29\\
 \noalign{\smallskip}\hline\noalign{\smallskip}
Children in the household & No children  & 0.41\\
& Children under 12 & 0.05\\
& Children over 12 & 0.54\\
\noalign{\smallskip}\hline\noalign{\smallskip}
 Main activity & Employed & 0.37\\
 &	Unemployed & 0.13\\
 &	Retired & 0.50\\
\noalign{\smallskip}\hline\noalign{\smallskip}
 \end{tabular}
\end{table}

\section{Mixture of latent trait analyzers}\label{sec:mlta}
Let $i=1,\dots,N$ denote the sending nodes and $k=1,\dots,R$ the receiving nodes. Bipartite network can be formally described by a random incidence matrix $\boldsymbol{Y} = \{Y_{ik}\}$, with elements
\begin{equation*} 
    Y_{ik}=\Bigg\{\begin{array}{@{}l@{}}
    1 \quad \mbox{ if sending node } i \mbox{ is connected with receiving node } k,\\
    0 \quad \mbox{ otherwise.}
  \end{array}\end{equation*}
The Mixture of Latent Trait Analyzers (MLTA; Gollini \& Murphy, 2014; Gollini, 2020) combines latent class and latent trait analysis by assuming that the set of $N$ sending nodes can be divided into $G$ distinct classes (or groups) and that the propensity of each sending node to be connected with the $R$ receiving nodes also depends on a multidimensional continuous latent trait. Therefore, the MLTA model first assumes that every sending node belongs to one of $G$ unobserved groups identified by the latent random variable $\boldsymbol{z}_i=(z_{i1}, \dots, z_{iG})^{\prime} \stackrel{iid}\sim \mbox{Multinomial}(1,(\eta_{1}, \dots, \eta_{G}))$, whose generic element is
 \begin{equation*}
    z_{ig}=\Bigg\{\begin{array}{@{}l@{}}
    1 \quad \mbox{ if sending node } i \mbox{ belongs to group } g,\\
    0 \quad \mbox{ otherwise.}
    \end{array}
\end{equation*}
The parameter $\eta_g$ denotes the probability that a randomly selected sending node belongs to group $g$, with $g = 1, \dots, G$, under the constraints that $\sum_{g=1}^G\eta_g=1$ and $\eta_g\geq0$, $g = 1, \dots ,G$.\\Furthermore, the model assumes the existence of a $D$-dimensional continuous latent trait $\boldsymbol{u}_i$ distributed according to a $D$-variate Gaussian density with null mean vector and identity covariance matrix, i.e. $\boldsymbol{u}_i \sim \mathcal{N}_D(\boldsymbol{0}, \boldsymbol{I})$, which captures the {heterogeneity} of the connections between sending and receiving nodes. Thus, response variables contained in the vector $\boldsymbol{y}_i=(y_{i1},\dots,y_{iR})^{\prime}$ are assumed to be independent Bernoulli random variables with parameters $\pi_{gk}(\boldsymbol{u}_i)$, $k = 1, \dots ,R$, modelled through the following logistic function:
\begin{equation}
\label{pi1}
    \pi_{gk}(\boldsymbol{u}_{i})=\text{Pr}(y_{ik}=1\mid \boldsymbol{u}_{i}, z_{ig}=1)=\frac{1}{1+\exp[-(b_{gk}+\boldsymbol{w}_{gk}^{\prime}\boldsymbol{u}_{i})]}.
\end{equation}
Here, the model intercept $b_{gk}$ represents the attractiveness of the $k$-th receiving node for sending nodes belonging to the $g$-th group. A positive and high value for this parameter indicates high attractiveness of the $k$-th receiving node for sending nodes in the $g$-th cluster, while a negative value indicates a low propensity of sending nodes in the $g$-th group to be connected to the $k$-th receiving node. Furthermore, the slopes $\boldsymbol{w}_{gk}$ associated with the latent variable $\boldsymbol{u}_{i}$ are meant to capture the influence of the latent trait on the probability of a connection between the $k$-th receiving node and the sending nodes belonging to the $g$-th group. Therefore, statistically significant estimates of $\boldsymbol{w}_{gk}$ indicate association between receiving nodes, as well as the presence of heterogeneity within groups with respect to the baseline level given by the intercept $b_{gk}$.\\Our contribution is to further extend the MLTA model to perform a joint clustering of sending and receiving nodes, also taking into account the effect of concomitant variables on the latent model structure, as described in the next section.

\section{Extension of MLTA in a multilevel setting}\label{sec:proposal}
In this section, the basic assumptions of the model and the parameter estimation method are presented.

\subsection{Model assumptions}
To adapt the model to the multilevel context of multi-layer bipartite networks, the notation introduced in Section 3 needs to be slightly modified. Let $n_h$ denote the number of sending nodes within layer $h$, with $h=1,\dots,H$. The relationship structure of a multi-layer bipartite network can be formally described by a random {incidence array} $\bm{Y} = \{Y_{hik}\}$, whose generic element is given by
\[ 
Y_{hik}=\Bigg\{\begin{array}{@{}l@{}}
    1 \quad \mbox{ if sending node } {i} \mbox{ within layer } h \mbox{ is connected with receiving node } k,\\
    0 \quad \mbox{ otherwise.}
  \end{array}
\]
The model assumes that sending nodes can be grouped into homogeneous classes identified by the latent random variable $\bm{z}_{hi}=(z_{hi1}, \dots, z_{hiG})^{\prime} \stackrel{iid}\sim \mbox{Multinomial}(1,(\eta_{hi1}, \dots, \eta_{hiG})^{\prime})$, whose generic element is given by
\[ 
z_{hig}=\Bigg\{\begin{array}{@{}l@{}}
    1 \quad \mbox{ if sending node } {i} \mbox{ within layer } h \mbox{ belongs to group } g,\\
    0 \quad \mbox{ otherwise.}
  \end{array}
\]
Here, $\eta_{hig}$ denotes the prior probability that the $i$-th sending node in the $h$-th layer belongs to group $g$, with $g = 1, \dots, G$. To account for the effect that nodal attributes may have on the clustering of nodes, we follow the approach of Failli et al. (2022) and include covariates in the latent layer of the model by employing a latent class regression model (Agresti, 2002). In addition, we also include a layer-specific random effect that captures unobserved sources of heterogeneity between layers:
\begin{equation}\label{eta2}
    \eta_{hig}=\text{Pr}(z_{hig}=1\mid \bm{x}_{hi}; \bm{\beta}_g, \epsilon_h)=\frac{\exp\{\bm{x}^{\prime}_{hi}\bm{\beta}_g+\epsilon_h\}}{1+\sum_{g'=2}^G\exp\{\bm{x}^{\prime}_{hi}\bm{\beta}_{g'}+\epsilon_h\}}, \quad g=2, \dots, G.
\end{equation}
Here, $\bm{x}_{hi}$ denote a $J$-dimensional vector of nodal attributes for sending node ${i}$ in layer $h$, $\bm{\beta}_g$ denote the corresponding $J$-dimensional vector of coefficients for the $g$-th latent class, and $\epsilon_h$, $h=1, \dots, H$, denote the layer-specific random effect. In this framework, a multilevel approach allows to properly account for the hierarchical structure of the data. {In detail, this allows not only to deal with unobserved sources of heterogeneity at multiple levels, but also to take into account the possible correlation between responses provided by units (sending nodes) belonging to the same hierarchical level.}\\To avoid unverifiable assumptions on the distribution of the random parameters $\epsilon_h$, we adopt a semi-parametric approach based on leaving such a distribution unspecified. Estimating this directly from the data, we obtain a discrete distribution with a finite number of support points $\{\gamma_1, \dots, \gamma_Q\}$ and weights $\{\rho_1, \dots, \rho_Q\}$. This is known as a non-parametric maximum likelihood (NPML; Laird, 1978) estimate of the mixing distribution on $\gamma_h$. According to this approach, the discrete distribution can be interpreted as representing subgroups in the data, such that each layer belongs to one of $Q$ unobserved groups identified by the latent random variable $v_{hq}$, whose generic element is
\[ 
v_{hq}=\Bigg\{\begin{array}{@{}l@{}}
    1 \quad \mbox{ if layer } {h} \mbox{ belongs to group } q,\\
    0 \quad \mbox{ otherwise,}
  \end{array}
\]
with $q=1,\dots,Q$. Here, $\rho_q =\text{Pr}(v_{hq} = 1)$ is the prior probability that the $h$-th layers belongs to group $q$, under the constraints that $\sum_{q=1}^Q \rho_q =1$ and $\rho_q\geq 0\; \forall \; q=1,\dots,Q$. Therefore, Equation (\ref{eta2}) becomes:
\[
\eta_{higq} = \text{Pr}(z_{hig} = 1 \mid \bm{x}_{hi}; \bm{\beta}_g, \gamma_{q}) = \frac{\exp\{\bm{x}^{\prime}_{hi}\bm{\beta}_g+\gamma_q\}}{1+\sum_{g'=2}^G\exp\{\bm{x}^{\prime}_{hi}\bm{\beta}_{g'}+\gamma_q\}}, \quad g=2, \dots, G,
\]
where, compared to the approach of Failli et al. (2022), membership probabilities may vary across layers' clusters.\\In addition, the propensity of network connections within each class also depends on the presence of a $D$-dimensional continuous latent trait $\bm{u}_{hi} \sim \mathcal{N}_D(\boldsymbol{0}, \boldsymbol{I})$, capturing the residual heterogeneity in the responses of units belonging to the $g$-th component, as in the latent trait framework. This means that the conditional distribution of the vector $\bm{y}_{hi} = (y_{hi1}, \dots, y_{hiR})^{\prime}$, given that node ${i}$ in layer $h$ belongs to the $g$-th group, is specified by a latent trait model with parameters $b_{gk}$ and $\bm{w}_{gk}$, $g = 1, \dots, G$, and $k = 1, \dots, R$. Thus, conditional on $\bm{z}_{hi}$ and $\bm{u}_{hi}$, variables contained in the $\bm{y}_{hi}$ vector are assumed to be independent Bernoulli random variables with parameters $\pi_{gk}(\bm{u}_{ig})$, $k = 1, \dots, R$, modelled via the following logistic function:
\begin{equation}
\label{pi}
    \pi_{gk}(\bm{u}_{hi})=\text{Pr}(y_{hik}=1\mid \bm{u}_{hi}, z_{hig}=1)=\frac{1}{1+\exp[-(b_{gk}+\bm{w}_{gk}^{\prime}\bm{u}_{hi})]}, \quad 0\leq \pi_{gk}(\bm{u}_{hi})\leq 1. 
\end{equation}
Here, the meaning and interpretation of parameters $b_{gk}$ and $\bm{w}_{gk}$ is the same as the original MLTA model introduced in Section 3.

\subsection{Parameter estimation}
Let $\boldsymbol{\theta}=(\boldsymbol{\beta}_2, \dots, \boldsymbol{\beta}_G, b_{11}, \dots, b_{GR}, \boldsymbol{w}_{11}, \dots, \boldsymbol{w}_{GR}, \gamma_1, \dots, \gamma_Q, \rho_1, \dots, \rho_{Q-1})$ be the vector of all free model parameters. Given the assumptions described in the previous section, the log-likelihood function of the model can be written as:
\begin{equation}
    \ell(\boldsymbol{\theta})=\sum_{h=1}^H \log \Big(\sum_{q=1}^Q \rho_q \prod_{i=1}^{n_h} \sum_{g=1}^G \eta_{higq} \int \prod_{k=1}^R f({y}_{hik}\mid \boldsymbol{u}_{hi}, z_{hig}=1)f(\boldsymbol{u}_{hi})d\boldsymbol{u}_{hi} \Big).
    \label{loglikelihood}
\end{equation}
where $f(y_{hik}\mid \boldsymbol{u}_{hi}, z_{hig} = 1) = (\pi_{gk}(\bm{u}_{hi}))^{y_{hik}}(1 - \pi_{gk}(\boldsymbol{u}_{hi}))^{1-y_{hik}}$. From Equation (4) we can easily recognise the log-likelihood function of a hierarchical mixture model.\\Maximum likelihood estimates of model parameters can be obtained by maximizing the log-likelihood function in Equation (4) via an EM algorithm (Dempster et al., 1977), which represents a standard estimation algorithm when dealing with latent variables. However, the integral to be solved in Equation (4) cannot be computed analytically as it does not admit a closed-form expression. Therefore, we propose a variational approximation of the log-likelihood function, following the approach of Jaakkola \& Jordan (1997). The proposed estimation algorithm is detailed in the next sections.

\subsubsection{Variational approximation}
The goal of the variational approach is to approximate the logistic function in Equation (3) with a lower bound that is conjugate with the Gaussian distribution assumed for $\boldsymbol{u}_{hi}$. Since the logistic function is convex, any of its tangents lies below the function and is part of the family of lower-bounds parameterized by their locations. Following an approach similar to that of Jaakkola \& Jordan (1997), the conditional distribution $f(y_{hik}\mid \boldsymbol{u}_{hi}, z_{hig} = 1)$ can be approximated by a function of auxiliary variational parameters $\boldsymbol{\xi}_{hig} = (\xi_{hig1}, \dots , \xi_{higR})^{\prime}$, with $\xi_{higk} \neq 0$, $\forall \; k = 1, \dots, R$, as follows
$$ \tilde{f}(\boldsymbol{y}_{hi}\mid \boldsymbol{u}_{hi}, z_{hig}=1, \boldsymbol{\xi}_{hig})=g(\xi_{higk})\exp\Big\{\frac{A_{higk}-\xi_{higk}}{2}+\lambda(\xi_{higk})(A_{higk}^2-\xi_{higk}^2)\Big\},$$
where:
\[A_{higk}=(2y_{hik}-1)(b_{gk}+\boldsymbol{w}_{gk}^{\prime}\boldsymbol{u}_{hi}),\]
\[g(\xi_{higk})=(1+e^{-\xi_{higk}})^{-1},\]
\[\lambda(\xi_{higk})=\frac{1/2-g(\xi_{higk})}{2\xi_{higk}}.\]
Thus, the logarithm of the component densities is approximated with a lower bound  which is conjugate with $f(\boldsymbol{u}_{hi})$, since it depends quadratically in the exponential on the latent trait $\boldsymbol{u}_{hi}$. After doing a little algebra, the logarithm of the integral in Equation (4) can be written as:
\begin{small}
\begin{align}\label{L}
    \mathcal{L}({\boldsymbol{\xi}}_{hig}) &= \log \tilde{f}(\boldsymbol{y}_{hi}\mid z_{hig}=1, \boldsymbol{\xi}_{hig}) \nonumber \\
    &=\log\int \tilde{f}(\boldsymbol{y}_{hi}\mid \boldsymbol{u}_{hi}, z_{hig}=1, \boldsymbol{\xi}_{hig}) f(\boldsymbol{u}_{hi}) d\boldsymbol{u}_{hi} \nonumber \\
    &= \log \int \prod_{k=1}^R g(\xi_{higk}) \exp\Big\{\Big(y_{hik}-\frac{1}{2}\Big)b_{gk} - \frac{\xi_{higk}}{2} + \lambda(\xi_{higk})b_{gk}^2 - \lambda(\xi_{higk})\xi_{higk}^2 + \frac{\boldsymbol{\mu}_{hig}^{\prime}\boldsymbol{\Sigma}_{hig}\boldsymbol{\mu}_{hig}}{2}\Big\} \nonumber \\ 
    & \mid\boldsymbol{\Sigma}_{hig}\mid^{1/2} \times {(2\pi)^{-D/2}\mid\boldsymbol{\Sigma}_{hig} \mid^{-1/2} \exp\Big\{-\frac{1}{2}(\boldsymbol{u}_{hi}-\boldsymbol{\mu}_{hig})\boldsymbol{\Sigma}_{hig}^{-1}(\boldsymbol{u}_{hi}-\boldsymbol{\mu}_{hig})^{\prime} \Big\}} d\boldsymbol{u}_{hi} .
\end{align}
\end{small}
In the last part of the above Equation, it is possible to recognise the kernel of a multivariate Gaussian density with covariance matrix and mean vector given by \[ \boldsymbol\Sigma_{hig} = \Big[\boldsymbol{I}-2\sum_k \lambda(\xi_{higk})\boldsymbol{w}_{gk}^{\prime}\boldsymbol{w}_{gk}\Big]^{-1}, \]
\[\boldsymbol\mu_{hig}=\boldsymbol\Sigma_{hig}\Big[\sum_k y_{hik}-\frac{1}{2} + 2\lambda(\xi_{higk})b_{gk}\Big]\boldsymbol{w}_{gk}.\]
Therefore, the integral in Equation (5) can be computed in closed-form, obtaining
\begin{align}\label{approx}
    \tilde{f}(\boldsymbol{y}_{hi}\mid z_{hig}=1, \boldsymbol{\xi}_{hig}) &= \exp\{ \mathcal{L}({\boldsymbol{\xi}}_{hig})\} \nonumber \\
    &=\exp \Big\{ \sum_k \Big[ \log g(\xi_{higk}) +  (y_{hik}-\frac{1}{2})b_{gk} - \frac{\xi_{higk}}{2} + \lambda(\xi_{higk})b_{gk}^2 - \lambda(\xi_{higk})\xi_{higk}^2  \Big]  \nonumber \\
    & + \frac{\boldsymbol{\mu}_{hig}^{\prime}\boldsymbol{\Sigma}_{hig}\boldsymbol{\mu}_{hig}}{2} + \frac{\log \mid\boldsymbol{\Sigma}_{hig}\mid}{2} \Big\}.
\end{align}
Thus, the log-likelihood is approximated as:
$$\tilde{\ell}(\boldsymbol{\theta})=\sum_{h=1}^H \log \Bigg(\sum_{q=1}^Q \rho_q \prod_{i=1}^{n_h} \sum_{g=1}^G \eta_{hiqg} \tilde{f}(\boldsymbol{y}_{hi}\mid z_{hig}=1, \boldsymbol{\xi}_{hig})\Bigg).$$

\subsubsection{EM algorithm}
The EM algorithm alternates two steps until convergence: at the E-step, the expected value of the complete data log-likelihood given the observed data and the current parameters value is computed; at the M-step, the expected complete data log-likelihood is maximized with respect to model parameters. These two steps are iterated until convergence is reached. In this framework, the complete data log-likelihood function is
\begin{align}\label{complete}
    \ell_c(\bm{\theta})=& \sum_{h=1}^H \sum_{i=1}^{n_h} \sum_{g=1}^G z_{hig} \log {f}(\boldsymbol{y}_{hi}\mid \boldsymbol{u}_{hi}, z_{hig}=1, \boldsymbol{\xi}_{hig}) \nonumber \\
    & + \sum_{h=1}^H \sum_{i=1}^{n_h} \log f(\boldsymbol{u}_{hi}) \nonumber \\
    & + \sum_{h=1}^H \sum_{q=1}^Q \sum_{i=1}^{n_h} \sum_{g=1}^G v_{hq}\; z_{hig} \; \log f(z_{hig}=1\mid v_{hq}=1) \nonumber \\
    & + \sum_{h=1}^H \sum_{q=1}^Q v_{hq} \log f(v_{hq}=1).
\end{align}
As it has been pointed out before, at the $(t + 1)$-th iteration of the algorithm, the E-step consists in computing the expectation of the complete data log-likelihood function
specified in Equation (7), conditional on the observed data and the current parameter estimates $\bm{\theta}^{(t)}$:
\begin{align}\label{Q}
    Q(\bm{\theta}\mid \bm{\theta}^{(t)}) & = \sum_{h=1}^H \sum_{i=1}^{n_h} \sum_{g=1}^G \int \hat{z}_{hig}^{(t+1)} \log {f}(\boldsymbol{y}_{hi}\mid \boldsymbol{u}_{hi}, z_{hig}=1, \boldsymbol{\xi}_{hig}) f(\boldsymbol{u}_{hi}) d\boldsymbol{u}_{hi} \nonumber \\
    & + \sum_{h=1}^H \sum_{i=1}^{n_h} \int f(\boldsymbol{u}_{hi}\mid \boldsymbol{y}_{hi}) \log f(\boldsymbol{u}_{hi}) d\boldsymbol{u}_{hi} \nonumber \\
    & + \sum_{h=1}^H \sum_{q=1}^Q \sum_{i=1}^{n_h} \sum_{g=1}^G \hat{a}_{higq}^{(t+1)}  \log (\eta_{higq})\nonumber\\
    & + \sum_{h=1}^H \sum_{q=1}^Q \hat{v}_{hq}^{(t+1)}\log (\rho_q). 
\end{align}
Here, $\hat{z}_{hig}^{(t+1)}$ denote the posterior expectation
of $z_{hig}$, given the current parameter estimates $\bm{\theta}^{(t)}$ and the observed data $\bm{y}_{hi}$, while $\hat{a}_{higq}^{(t+1)}$ denote the joint posterior expectation
of $z_{hig}$ and $v_{hq}$, given the current parameter estimates $\bm{\theta}^{(t)}$ and the observed data $\bm{y}_{hi}$, and $\hat{v}_{hq}^{(t+1)}$ denote the posterior expectation of $v_{hq}$, given the current parameter estimates $\bm{\theta}^{(t)}$ and the observed data $\bm{y}_{hi}$. Removing the dependence on the iteration index and doing a little algebra, posterior probabilities can be computed as
$$ \hat{z}_{hig} = \frac{\sum_{q=1}^{Q} \rho_q \; \eta_{higq}\;\tilde{f}(\boldsymbol{y}_{hi}\mid z_{hig}=1, \boldsymbol{\xi}_{hig})}{\sum_{q=1}^Q \rho_q \sum_{g=1}^G \eta_{hiqg}\; \tilde{f}(\boldsymbol{y}_{hi}\mid z_{hig}=1, \boldsymbol{\xi}_{hig})}, $$
\vspace{-7mm}
\begin{align}\label{vhq}
    \hat{v}_{hq} = \frac{\rho_q \prod_{i=1}^{n_h}\sum_{g=1}^G \eta_{hiqg} \;\tilde{f}(\boldsymbol{y}_{hi}\mid z_{hig}=1, \boldsymbol{\xi}_{hig})}{\sum_{q=1}^Q \rho_q \prod_{i=1}^{n_h}\sum_{g=1}^G \eta_{hiqg}\; \tilde{f}(\boldsymbol{y}_{hi}\mid z_{hig}=1, \boldsymbol{\xi}_{hig})},
\end{align}
where $\tilde{f}(\boldsymbol{y}_{hi}\mid z_{hig}=1, \boldsymbol{\xi}_{hig})$ is the closed-form solution of the integral in Equation (4) obtained through the variational approach (see Equation (6)).\\The joint posterior expectation of $z_{hig}$ and $v_{hq}$ can be obtained by exploiting the conditional independence assumption (Vermunt \& Magidson, 2005):
\begin{align}
    \hat{a}_{higq}=f(v_{hq}=1,z_{hig}=1\mid \bm{y}_h)=f(v_{hq}=1\mid \boldsymbol{y}_h) f(z_{hig}=1 \mid \boldsymbol{y}_{hi}, v_{hq}=1),
\end{align}
where:
\begin{align}
    f(z_{hig}=1\mid \boldsymbol{y}_{hi}, v_{hq}=1) = \frac{\eta_{hiqg}\; \tilde{f}(\boldsymbol{y}_{hi}\mid z_{hig}=1, \boldsymbol{\xi}_{hig}) }{\sum_{g=1}^G \eta_{hiqg} \; \tilde{f}(\boldsymbol{y}_{hi}\mid z_{hig}=1, \boldsymbol{\xi}_{hig}) },
\end{align}
and $f(v_{hq}=1\mid \boldsymbol{y}_h)$ is computed as in Equation (9).\\The M-step of the algorithm consists in updating the model parameters by maximizing the expected complete data log-likelihood function in Equation (8) with respect to $\boldsymbol{\theta}$. To this aim, the following steps are carried out:
\begin{itemize}
    \item[1)] The multinomial logit coefficients $\boldsymbol\beta_g$ and $\gamma_q$ are estimated via a Newton-Raphson algorithm maximizing the likelihood of a multinomial logit model, with weights provided by $\hat{a}_{higq}$ obtained at the previous step. Prior probabilities $\eta_{hiqg}$ are updated accordingly. Moreover, $\rho_q$ is updated as:
    $$\hat{\rho}_q=\frac{\sum_{h=1}^H f(v_{hq}=1\mid \boldsymbol y_h)}{H}.$$
    \item[2)] A second EM algorithm nested within the first is required to update the posterior mean vector and covariance matrix of $\boldsymbol{u}_{hi}$, the variational parameters $\xi_{higk}$, and the latent trait parameters $b_{gk}$ and $\boldsymbol{w}_{gk}$. These are obtained as follows:
    \begin{itemize}
        \item[a)]  In the $(t + 1)$-th iteration of the E-step, we compute the approximate posterior distribution of the continuous latent variable $\bm{u}_{hi}$, given the
observation $\bm{y}_{hi}$, the posterior probability $\hat{z}_{hig}^{(t+1)}$, and the variational parameters $\hat{\bm{\xi}}_{hig}^{(t)}$. The posterior density of $\boldsymbol{u}_{hi}$ is obtained as:
        \begin{align*}
            \tilde{f}(\boldsymbol{u}_{hi}\mid \boldsymbol{y}_{hi}, z_{hig}=1, \boldsymbol{\xi}_{hig}) &= \frac{\tilde{f}(\boldsymbol{y}_{hi}\mid z_{hig}=1, \boldsymbol{u}_{hi}, \boldsymbol{\xi}_{hig})\; f(\boldsymbol{u}_{hi})}{\int \tilde{f}(\boldsymbol{y}_{hi}\mid z_{hig}=1, \boldsymbol{u}_{hi}, \boldsymbol{\xi}_{hig})\; f(\boldsymbol{u}_{hi}) d\boldsymbol{u}_{hi}} \nonumber \\    
             &= \prod_{k} g(\xi_{higk}) (2\pi)^{-D/2} \mid \boldsymbol{\Sigma}_{hig}\mid ^{-1/2} \times \nonumber \\
             &\exp \Big\{ \Big(y_{hik}-\frac{1}{2}\Big)b_{gk} -\frac{\xi_{higk}}{2} + \lambda(\xi_{higk})b_{gk}^2 -\lambda(\xi_{higk})\xi_{higk}^2  \Big\} \times \nonumber \\
             &\exp\Big\{- \frac{1}{2}(\boldsymbol{u}_{hi}-\boldsymbol{\mu}_{hig})\boldsymbol{\Sigma}_{hig}^{-1}(\boldsymbol{u}_{hi}-\boldsymbol{\mu}_{hig})^{\prime}\Big\} \times \nonumber \\
             &\exp\Big\{ \frac{\boldsymbol{\mu}_{hig}^{\prime}\boldsymbol{\Sigma}_{hig}^{-1}\boldsymbol{\mu}_{hig}}{2}\Big\} \mid \boldsymbol{\Sigma}_{hig}\mid^{1/2}
            \end{align*}
        Here, we recognize the kernel of a gaussian distribution with covariance matrix $\boldsymbol\Sigma_{hig}$ and mean vector $\boldsymbol\mu_{hig}$, given by:
        \[ \boldsymbol\Sigma_{hig} = \Big[\mathbb{I}-2\sum_k \lambda(\xi_{higk})\boldsymbol{w}_{gk}^{\prime}\boldsymbol{w}_{gk}\Big]^{-1}, \]
        \vspace{-10mm}
        \[\boldsymbol\mu_{hig}=\boldsymbol\Sigma_{hig} \Big[\sum_k y_{hik}-\frac{1}{2} + 2\lambda(\xi_{higk})b_{gk}\Big]\boldsymbol{w}_{gk}.\]
        \item[b)]In the M-step, the variational parameters $\xi_{higk}$ are updated by maximizing the first part of Equation (8). The first derivative with respect to $\xi_{higk}$ is
        \[\frac{\partial\lambda(\xi_{higk})}{\partial \xi_{higk}}\mathbb{E}[(b_{gk}+\boldsymbol{w}_{gk}^{\prime}\boldsymbol{u}_{hi})^2]-\xi_{higk}^2.\]
        Since $\lambda(\xi_{higk})$ is an increasing monotone function in $\xi_{higk}$, the maximization consists in:
        $$ \mathbb{E}[(b_{gk}+\boldsymbol{w}_{gk}^{\prime}\boldsymbol{u}_{hi})^2]-\xi_{higk}^2 =0, $$
        obtaining
        $$ \xi_{higk}^2 = b_{gk}^2 + 2 b_{gk}\boldsymbol{w}_{gk}^{\prime}\boldsymbol{\mu}_{hig} + \boldsymbol{w}_{gk}^{\prime}(\boldsymbol\Sigma_{hig} + \boldsymbol\mu_{hig}^{\prime}\boldsymbol\mu_{hig})\boldsymbol{w}_{gk}.  $$
    \item[c)] {Let $\hat{\bm{\zeta}}_{gk}^{(t+1)}=(\hat{\bm{w}}_{gk}^{(t+1)\prime},\hat{b}_{gk}^{(t+1)})^{\prime}$; updates for these parameters are obtained by maximizing the expected value of the complete data log-likelihood function, resulting in:}
\[
    \hat{\bm{\zeta}}_{gk}^{(t+1)}=-\Big[2\sum_{h=1}^{H}\sum_{i=1}^{n_h} \hat{z}_{hig}^{(t+1)}\lambda(\hat{\xi}_{higk}^{(t+1)})\mathbb{E}[\hat{\bm{u}}_{hi}\hat{\bm{u}}_{hi}^{\prime}]_g^{(t+1)}\Big]^{-1}\Big[\sum_{h=1}^H\sum_{i=1}^{n_h} \hat{z}_{hig}^{(t+1)}\Big(y_{hik}-\frac{1}{2}\Big)\hat{\bm{\alpha}}_{hig}^{(t+1)}\Big], 
\]
where 
\[
\hat{\bm{\alpha}}_{hig}^{(t+1)}=(\hat{\bm{\mu}}_{hig}^{(t+1)\prime}, 1)^\prime
\]
\[\mathbb{E}[\hat{\bm{u}}_{hi}\hat{\bm{u}}_{hi}^{\prime}]_g^{(t+1)}=
\left[
  \begin{array}{ c c }
    \hat{\bm{\Sigma}}_{hig}^{(t+1)}+\hat{\bm{\mu}}_{hig}^{(t+1)}\hat{\bm{\mu}}_{hig}^{(t+1)\prime} & \hat{\bm{\mu}}_{hig}^{(t+1)}\\
\hat{\bm{\mu}}_{hig}^{(t+1)\prime} & 1\\
  \end{array} \right].
\]
        \item[d)] The lower bound is then approximated as:
        \begin{align*}
     \tilde{f}(\bm{y}_{hi}\mid z_{hig}=1, \bm{\xi}_{hig}) & = \sum_{k=1}^R \Big[\log(g({\xi}_{higk}))-\frac{\hat{\xi}_{higk}}{2} -\lambda({\xi}_{higk}){\xi}_{higk}^{2}+\Big(y_{hik}-\frac{1}{2}\Big){b}_{gk} \nonumber \\
     &+ \lambda({\xi}_{higk}){b}_{gk}^{2}\Big]+\frac{\log|\hat{\boldsymbol{\Sigma}}_{hig}|}{2} + \frac{\hat{\boldsymbol{\mu}}_{hig}^{\prime}[\hat{\boldsymbol{\Sigma}}_{hig}]^{-1}\hat{\boldsymbol{\mu}}_{hig}}{2}.\nonumber
\end{align*}
    \end{itemize}
\end{itemize}
As typically happens with latent variables, an initialization strategy based on multiple starting points may help the estimation algorithm avoid being trapped in local maxima of the log-likelihood function.\\
When convergence of the algorithm is achieved, each sending node can be assigned to the $g$-th group via a Maximum a Posteriori (MAP) rule on the basis of the estimated posterior probabilities $\hat{z}_{hig}$, while each layer can be assigned to the $q$-th group according to the estimated posterior probabilities $\hat{v}_{hq}$.

\subsection{Standard errors and model selection}
\label{subsec:stderr}
To evaluate uncertainty associated with the estimates obtained from the EM algorithm, we employ a {non-parametric bootstrap} approach (Efron, 1979) within each layers, in order to account for the multilevel structure of the data (Goldstain, 2011). Given an incidence matrix $\bm{Y}$, the non-parametric bootstrap consists in repeatedly drawing with replacement a sample having the same size of $\bm{Y}$, and then estimating the model. In detail, for each bootstrap replicate, $n_h$ rows of the incidence matrix are drawn with repetition $\forall \; h=1,\dots,H$, so that each sending node within each layer can appear several times, and the model is estimated for fixed $G$ (number of sending nodes groups), $D$ (latent trait size) and $Q$ (number of layers groups).\\Let $\hat{\bm{\theta}}_{(s)}$ denote the vector of estimates obtained from the $s$-th bootstrap sample. By paying attention to the label switching issue, bootstrap standard errors correspond to the square root of the diagonal elements of the following matrix:
\[
    \mbox{V}(\hat{\bm{\theta}})=\frac{1}{S}\sum_{s=1}^S \big(\hat{\bm{\theta}}_{(s)}-\hat{\bm{\theta}}_{(.)}\big)\big(\hat{\bm{\theta}}_{(s)}-\hat{\bm{\theta}}_{(.)}\big)^{\prime},
\]
with $\hat{\bm{\theta}}_{(.)}$ being the empirical mean vector $\hat{\bm{\theta}}_{(.)}=\frac{1}{S}\sum_{s=1}^S \hat{\bm{\theta}}_{(s)}.$\\
Regarding model selection, the MLTA model is estimated for different values of the number of latent classes $G$, the size of the continuous latent variable $D$, and the number of layers groups $Q$. The model corresponding to the smallest value of the chosen information criterion, such as the {Bayesian Information Criterion} (BIC; Schwarz, 1978), is selected as the optimal one. 

\section{Simulation study}\label{sec:simulation}
To evaluate the ability of the proposal in correctly identifying model parameters and classifying sending nodes (and layers), we conduct a simulation study as described below.

\subsection{Simulation setup}
We consider different scenarios based on different number of sending nodes ($N$=500, 1000, 2000), receiving nodes ($R$=7, 14), and layers' groups ($Q$=2, 3), while the latent trait size $D$, the number of layers $H$, and the number of groups $G$, are kept constant and equal to $D=$1, $H$=20, and $G$=3, respectively. Furthermore, class membership is defined in terms of a single nodal attribute (i.e., $J$ = 1), which is drawn from a Gaussian distribution with mean and variance both equal to 1. The latent structure is defined by setting $\bm{\beta}_2 = (1, -0.4)$ and $\bm{\beta}_3 = (1.5, -0.9)$. Furthermore, we set $b_{gk}$ by simulating $\bm{b}_g = (b_{g1}, \dots, b_{gR})$ from a multivariate Gaussian distribution with mean vector $-3\bm{1}_R$, $0\bm{1}_R$, and $3\bm{1}_R$, for $g$ = 1, 2, and 3, respectively. The covariance matrix considered to simulate the $\bm{b}_g$ parameters is set to $\bm{I}$. As regards the $\bm{w}_{gk}$ parameters, we consider a more parsimonious version of the model by assuming constant $\bm{w}_{gk}$ parameters across groups (i.e., $\bm{w}_{gk} = \bm{w}_k \; \forall \; g = 1, \dots, G)$, meaning that the latent trait has the same effect in all groups. These are set by simulating from a standard Gaussian distribution. Finally, we set $\bm\gamma=(-0.5, 1.5)$ and $\bm\rho=(0.3, 0.7)$ when $Q$=2, and $\bm\gamma=(-0.5, 1.5, 2.5)$ and $\bm\rho=(0.33, 0.33, 0.33)$ when $Q$=3.\\
In each scenario, the number of random starting values for the model parameters is set equal to 100. Simulation results are based on $B$ = 100 samples.

\subsection{Simulation results: clustering recovery}
The ability of the proposal in correctly classifying sending nodes and layers is evaluated via the Adjusted Rand Index (ARI; \citealp{rand}), which measures the agreement between the true partitions and the estimated ones. Thus, its expected value is zero in the case of random partition, while it is 1 in the case of perfect agreement between the two partitions. Results of the simulation study are shown in Table 3. Looking at this table, we note that, as the size of the network increases (i.e., $N$ and $R$ increases), the classification improves for both sending nodes and layers. However, ARI values for sending nodes' partitions remain stable as the number of layers' partitions increases. On the other hand, the ARI values for layers' partitions decrease as the number of such partitions increases from $Q$=2 to $Q$=3. However, the classification ability of the model is expected to further improve when considering bipartite networks with a larger number of nodes, as in the real-data application discussed in Section 2.

\begin{table}[ht!]
\normalsize
\caption{Adjusted Rand Index mean (median) across samples for sending nodes and layers partitions with varying $N$, $R$, $G$ and $Q$.}
\label{tab:1}       
\centering
\scalebox{0.8}{
\begin{tabular}{p{1.4cm}cc}
\noalign{\smallskip}\hline\noalign{\smallskip}
 & \multicolumn{2}{c}{{Sending nodes}} \\
\cmidrule(lr){2-3} %
& \multicolumn{2}{c}{\textit{G}=3} \\
\cmidrule(lr){2-3} %
{\textit{R}=7} & {\textit{Q}=2} &  {\textit{Q}=3}  \\
\noalign{\smallskip}\hline\noalign{\smallskip}
{\textit{N}=}500 & 0.749 (0.753) & 0.751 (0.757) \\
{\textit{N}=}1000 & 0.758 (0.758) & 0.761 (0.761) \\
{\textit{N}=}2000 &  0.761 (0.761) & 0.765 (0.765) \\
\noalign{\smallskip}\hline\noalign{\smallskip}
& \multicolumn{2}{c}{\textit{G}=3} \\
\cmidrule(lr){2-3} %
{\textit{R}=14} & {\textit{Q}=2} &  {\textit{Q}=3} \\
\noalign{\smallskip}\hline\noalign{\smallskip}
{\textit{N}=}500 & 0.893 (0.896) & 0.886 (0.884) \\
{\textit{N}=}1000 & 0.891 (0.893) & 0.897 (0.897) \\
{\textit{N}=}2000 & 0.896 (0.896) & 0.894 (0.895) \\
\noalign{\smallskip}\hline\noalign{\smallskip}

 & \multicolumn{2}{c}{{Layers}} \\
\cmidrule(lr){2-3} %
& \multicolumn{2}{c}{\textit{G}=3} \\
\cmidrule(lr){2-3} %
{\textit{R}=7} & {\textit{Q}=2} &  {\textit{Q}=3} \\
\noalign{\smallskip}\hline\noalign{\smallskip}
{\textit{N}=}500 & 0.689 (0.775) & 0.396 (0.370) \\
{\textit{N}=}1000 & 0.897 (1.000) & 0.549 (0.553) \\
{\textit{N}=}2000 & 0.993 (1.000) & 0.597 (0.603) \\
\noalign{\smallskip}\hline\noalign{\smallskip}
& \multicolumn{2}{c}{\textit{G}=3}\\
\cmidrule(lr){2-3} %
{\textit{R}=14} & {\textit{Q}=2} &  {\textit{Q}=3} \\
\noalign{\smallskip}\hline\noalign{\smallskip}
{\textit{N}=}500 & 0.828 (0.800) & 0.458 (0.449) \\
{\textit{N}=}1000 & 0.947 (1.000) & 0.590 (0.596) \\
{\textit{N}=}2000 & 1.000 (1.000) & 0.665 (0.669) \\
\noalign{\smallskip}\hline\noalign{\smallskip}
\end{tabular}}
\end{table}

\subsection{Simulation results: model parameters}
Table 4 shows the Mean Squared Error (MSE) values across samples for $\bm{\beta}=(\bm{\beta}_1,\dots,\bm{\beta}_G)$, $\bm{\gamma}=(\gamma_1,\dots,\gamma_Q)$, $\bm{\rho}=(\rho_1,\dots,\rho_Q)$ by varying the number of nodes and partitions.\\As it is well known, the MSE is computed as the mean of squared differences between the estimates and the true value of model parameters. Therefore, the smaller the MSE, the closer the estimates and the true parameters are. Looking at Table 4, it is evident that, as the size of the network increases, we are more and more able to identify the true values of model parameters. However, MSE values increase as the number of partitions increases. In detail, for $Q$=3, larger network sizes are needed to obtain good results in terms of MSE. Again, we expect that by further increasing the number of sending and receiving nodes, estimates will improve.
\begin{table}[ht!]
\normalsize
 \caption{MSE values across samples for $\bm{\beta}$, $\bm{\gamma}$ and $\bm{\rho}$, with varying $N$, $R$, and $Q$.}
    \label{tab:MSE1}
    \centering
    \scalebox{0.63}{
    \begin{tabular}{l c c c}
    \toprule
      &  & \multicolumn{2}{c}{$G$=3} \\
\cmidrule(lr){3-4} %
\multicolumn{2}{c}{\textit{R}=7} & {\textit{Q}=2} &  {\textit{Q}=3} \\
     \midrule
     \multirow{3}{*}{$N$=500} & $\bm{\beta}$ & 0.238 0.233 0.047 0.034 & 0.567 0.599 0.051 0.042 \\
     & $\bm{\gamma}$ & 0.000 0.192 & 0.000 0.271 0.387 \\
     & $\bm{\rho}$ & 0.011 0.011 & 0.048 0.004 0.049 \\
     \midrule
     \multirow{3}{*}{$N$=1000} & $\bm{\beta}$ & 0.221 0.223 0.018 0.016 & 0.519 0.557 0.025 0.022 \\
     & $\bm{\gamma}$ & 0.000 0.073 & 0.000 0.114 0.280 \\
     & $\bm{\rho}$ & 0.010 0.010 & 0.044 0.005 0.043 \\
     \midrule
     \multirow{3}{*}{$N$=2000} & $\bm{\beta}$ & 0.259 0.305 0.009 0.009 & 0.607 0.658 0.012 0.013 \\
     & $\bm{\gamma}$ &  0.000 0.056 & 0.000 0.068 0.438 \\
     & $\bm{\rho}$ & 0.010 0.010 & 0.059 0.007 0.059 \\
     \toprule
      &  & \multicolumn{2}{c}{$G$=3} \\
\cmidrule(lr){3-4} %
\multicolumn{2}{c}{\textit{R}=14} & {\textit{Q}=2} &  {\textit{Q}=3}  \\
\midrule
     \multirow{3}{*}{$N$=500} & $\bm{\beta}$ & 0.269 0.259 0.027 0.024 & 0.337 0.356 0.028 0.027 \\
     & $\bm{\gamma}$ & 0.000 0.158 & 0.000 0.176 1.194 \\
     & $\bm{\rho}$ & 0.010 0.010 & 0.027 0.004 0.036 \\
     \midrule
     \multirow{3}{*}{$N$=1000} & $\bm{\beta}$ & 0.242 0.218 0.011 0.01 & 0.404 0.410 0.013 0.013 \\
     & $\bm{\gamma}$ & 0.000 0.114 & 0.000 0.090 0.173 \\
     & $\bm{\rho}$ & 0.010 0.010 & 0.024 0.004 0.032 \\
     \midrule
     \multirow{3}{*}{$N$=2000} & $\bm{\beta}$ & 0.270 0.274 0.005 0.007  & 0.309 0.310 0.007 0.008 \\
     & $\bm{\gamma}$ & 0.000 0.050 & 0.000 0.151 0.400 \\
     & $\bm{\rho}$ & 0.011 0.011 & 0.021 0.003 0.026 \\
     \bottomrule
    \end{tabular}}
\end{table}

\section{Results}\label{analysis}
As previously pointed out, we aim at analyzing the digital divide in Europe via the proposed modeling approach, which allows us to perform a clustering of respondents and countries, also taking into account how socio-economic and demographic features influence the propensity of being digitally skilled. To this aim, the model is estimated for varying number of groups $G$, latent trait size $D$, and number of country groups $Q$. To prevent the estimation algorithm from remaining trapped in local maxima, a multi-start strategy based on 10 random starts is considered. As regards the concomitant variables affecting the latent layer of the model, the categorical variables described in Table 2 have been considered.\\Table 5 reports the BIC values for all the estimated models; the optimal specification is the one providing the smallest BIC value corresponding to $G = 4$ groups, $D=1$ latent trait size, and $Q=2$ country groups.
\begin{table}[ht!]
\small
    \centering
\caption{BIC for varying numbers of components $G$ and segments $D$.}
    \begin{tabular}{l c c c c c}
    \noalign{\smallskip}\hline\noalign{\smallskip}
     $Q=1$ & $G=1$ & $G=2$ & $G=3$ & $G=4$ & $G=5$ \\
     \noalign{\smallskip}\hline\noalign{\smallskip}
$D=1$ & 93634.88 & 88459.42 & 86241.20 & 87033.70 & 93409.04  \\
$D=2$ & 91400.51 & 88557.90 & 85391.04 & 88472.25 & 99293.92  \\ 
$D=3$ & 91447.96 & 88539.75 & 86425.04 & 91238.00 & 90875.28  \\
\noalign{\smallskip}\hline\noalign{\smallskip}
$Q=2$ & $G=1$ & $G=2$ & $G=3$ & $G=4$ & $G=5$ \\
     \noalign{\smallskip}\hline\noalign{\smallskip}
$D=1$ & 93634.88 & 84126.13 & 82783.09 & \textbf{81848.99} & 82239.38 \\
$D=2$ & 91400.51 & 84135.27 & 83029.45 & 82009.49 & 81925.86\\ 
$D=3$ & 91447.96 & 84341.12 & 82945.94 & 81995.58 & 82904.78 \\
\noalign{\smallskip}\hline\noalign{\smallskip}
$Q=3$ & $G=1$ & $G=2$ & $G=3$ & $G=4$ & $G=5$ \\
     \noalign{\smallskip}\hline\noalign{\smallskip}
$D=1$ & 93634.88 & 83808.40 & 82803.18 & 81897.24 & 83575.87 \\
$D=2$ & 91400.51 & 83929.99 & 82839.41 & 83425.37 & 81948.35 \\ 
$D=3$ & 91447.95 & 84082.13 & 82974.73 & 82204.79 & 81858.22 \\
     \noalign{\smallskip}\hline\noalign{\smallskip}
    \end{tabular}
    \label{tab:bic}
\end{table}
The model groups Belgium, Switzerland, Czechia, Estonia, Finland, France, Ireland, Iceland, Netherlands, and Norway in the same cluster, while Bulgaria, United Kingdom, Croatia, Hungary, Italy, Lithuania, Montenegro, North Macedonia, Portugal, Slovenia, and Slovakia are assigned to the other group. Comparing these results with the Digital Economy and Society Index (DESI) index (European Commission, 2022), which measures EU Member States’ progress in digitalization, it can be seen that the first group includes the most digitalized countries (i.e., countries with a DESI index above 50\%), while the other comprises the least digitalized ones (i.e., countries with a DESI index below 50\%). Therefore, the proposed method is effective in clustering both individuals and countries and takes into account not only residual heterogeneity, but also the multilevel structure of the data.\\Table 6 shows the estimated ${b}_{gk}$ and $\bm{w}_k$ parameters with 95\% bootstrap confidence intervals. In this context, $b_{gk}$ represents the attractiveness of each digital skill for older adults belonging to the $g$-th group, while $\bm{w}_k$ capture the influence of the latent trait on the probability of having each digital skill. Furthermore, the latent trait can be interpreted as the unobserved propensity of each individual to possess a specific digital skill. Looking at Table 6, we observe that the estimates of the attractiveness coefficient in the first group for each digital skill, i.e. $\bm{b}_{1}$, are all negative and statistically different from zero, which means that individuals in the first group are characterised by a negative propensity to have all the different digital skills, {which means that group 1 tend to attract individuals who do not have any of the considered digital skills.} A similar situation is that of group 2, where the only positive and significant coefficient is the one related to message exchange. On the other hand, group 3 mainly differs from group 2 in that the former uses Internet in addition to exchanging messages. Finally, group 4 is characterised by an intensive use of all digital skills, with the exception of video calling and sharing online political posts, for which there is a lower propensity as evidenced by the corresponding negative coefficients. Therefore, groups can be divided according to the digitalization level, which is lower in the first and second groups, medium in the third group and higher in the fourth group. Finally, regarding the influence coefficient $\bm{w}$, we can see that individuals show homogeneity with respect to the use of advanced search tools (i.e., its corresponding coefficient is not statistically significant).
\begin{table}[ht!]
\normalsize
    \centering
\caption{Estimated ${b}_{gk}$ and $\bm{w}_k$ parameters with 95\% bootstrap CIs.}
\scalebox{0.67}{
    \begin{tabular}{l c c c c c}
    \noalign{\smallskip}\hline\noalign{\smallskip}
     & $\bm{b}_1$ & $\bm{b}_2$ & $\bm{b}_3$ & $\bm{b}_4$ & $\bm{w}$\\
         \noalign{\smallskip}\hline\noalign{\smallskip}
      Internet & -2.33 (-2.63; -2.04)  & -1.67 (-3.2; -0.31) &  1.94 (1.58; 2.30) & 3.24 (3.11; 3.38) &  0.12 (0.09; 0.16) \\
      Preference setting & -5.31 (-6.11; -4.50) & -4.46 (-5.50; -3.42) & -1.40 (-1.69; -1.10) & 2.44 (2.31; 2.56)  & 0.04 (0.01; 0.07)\\
      Advanced search & -6.20 (-7.30; -5.10) & -4.86 (-5.78; -3.93) & -1.21 (-1.54; -0.88) & 3.29 (3.07; 3.50) & 0.01 (-0.02; 0.03)\\
      PDF & -6.56 (-7.91; -5.22) & -5.28 (-5.98; -4.58) & -1.05 (-1.31; -0.79) & 2.15 (2.06; 2.25)  & 0.05 (0.03; 0.07)\\
      Video call &  -13.66 (-15.75; -11.57) & 0.19 (-0.02; 0.40) & -0.76 (-0.93; -0.59) & -0.57 (-0.67; -0.48) & 1.09 (1.04; 1.13)\\
      Messages & -13.14 (-16.67; -9.61) & 0.60 (0.41; 0.78) & 0.20 (0.04; 0.36) & 0.52 (0.34; 0.71) &  1.10 (1.05; 1.15)\\
      Online posts & -7.76 (-8.67; -6.85) & -5.60 (-6.21; -4.98) & -2.17 (-2.35; -1.99) & -1.42 (-1.48; -1.35) & 0.14 (0.10; 0.19)\\
     \noalign{\smallskip}\hline\noalign{\smallskip}
    \end{tabular}}
    \label{tab:bw}
\end{table}\\
To better interpret results, Table 7 shows the predicted probabilities of having each digital skill for individuals belonging to each group, computed using Equation (3). It is evident from this table that the first group is characterised by very low probabilities and almost zero for each of the digital skills considered in the analysis. In the second group, on the other hand, the probability of adopting a certain skill is only high for the use of video calls and messages. On the contrary, in the third group, the probability tends to be higher and the probability of using the Internet and messaging reaches 89\% and 79\%, respectively. Finally, the fourth group is characterised by a high probability of possessing each of the skills considered, with the exception of sharing political posts online, for which, anyway, the predicted probability is the highest across all groups. The second group deserves special attention, since the probability of using the Internet is only 18\% while that of making video calls is 78\%. From this result, we can assume that the second group consists of people who declare they use the Internet rarely or never, but if they do, it is only for video calls and messages. 
\begin{table}[ht!]
    \centering
    \footnotesize
     \caption{Predicted probabilities of having each digital skill for individuals belonging to each group.}
    \begin{tabular}{{p{1.1cm}p{1.1cm}p{1.1cm}p{1.1cm}p{1.1cm}p{1.1cm}p{1.1cm}p{1.1cm}}}
    \toprule
         &  \multicolumn{7}{c}{Digital skills}\\
      \cmidrule(lr){2-8}
       Group  & Internet use & Preference settings & Advanced search & PDFs & Video calls & Messages & Online posts\\
       \midrule
         $G$=1 & 0.10 & 0.01 & 0.00 & 0.00 & 0.00 & 0.00 & 0.00\\
         $G$=2 & 0.18 & 0.01 & 0.01 & 0.01 & 0.78 & 0.84 & 0.00\\
         $G$=3 & 0.89 & 0.20 & 0.23 & 0.27 & 0.58 & 0.79 & 0.12\\
         $G$=4 & 0.97 & 0.92 & 0.96 & 0.90 & 0.63 & 0.83 & 0.22\\
         \bottomrule
    \end{tabular}
    \label{prob}
\end{table}\\Table 8 shows estimates and 95\% bootstrap confidence intervals of the $\bm{\beta}_g$ parameters, i.e. the effect of demographic and socio-economic variables on the probability of belonging to group $g=2,\dots,G$. In this context, the reference group is the first, which represents the group of less digitalized individuals. Therefore, Table 8 shows that the probability of belonging to group 2, compared to the probability of belonging to group 1, is higher for individuals who are between 65 and 75 years of age, healthier, hampered in daily activities, medium/low educated, with medium-low income, with young children or without children, and for retired or unemployed individuals. On the other hand, the probability of belonging to the medium digitalized group (group 3), is higher for individuals between 65 and 75 years of age, in good health, hampered in daily activities, with low income, medium educated, with a low/medium educated partner, with an average income, with young children or without children and for unemployed respondents. On the other hand, this probability is lower for those over 75, with a low level of education, low income and retired. Finally, the likelihood of belonging to the most digitalized group (group 4) is higher for those who have a good level of health, are hampered in daily activities, were born in the country of residence, are male, have a high level of education, have a partner, have a medium level of income, have young children or no children and are unemployed. On the contrary, this probability is lower for those who are over 75 years old, have a low income and are retired.
\begin{table}[ht!]
\footnotesize
    \centering
\caption{Estimated $\bm{\beta}_g$ parameters and 95\% bootstrap CIs. The reference category is in brackets.}
    \begin{tabular}{l c c c}
    \noalign{\smallskip}\hline\noalign{\smallskip}
     & $\bm{\beta}_2$ & $\bm{\beta}_3$ & $\bm{\beta}_4$\\
         \noalign{\smallskip}\hline\noalign{\smallskip}
      Intercept & -1.26 (-1.32; -1.20) &  -0.01 (-0.06; 0.05) & 0.91 (0.87; 0.94)\\
      Age 65-75 (50-65) & 0.47 (0.36; 0.58) & 0.28 (0.16; 0.40) & 0.10 (-0.05; 0.25)\\
      Age 75+ (50-65) & 0.28 (-0.12; 0.68) &  -0.32 (-0.63; -0.02) & -0.86 (-1.17; -0.54)\\
      Health fair (bad) & 0.12 (0.02; 0.22) & 0.18 (0.12; 0.24) & 0.47 (0.35; 0.59)\\ 
      Health good (bad) & 0.26 (-0.10; 0.62) &  0.64 (0.56; 0.72) & 1.41 (1.26; 1.57)\\
      Not hampered & 0.28 (0.09; 0.47) & 0.16 (0.11; 0.22) & 0.26 (0.21; 0.32)\\
      Born in country & -0.01 (-0.06; 0.04) & 0.01 (-0.05; 0.07) & 0.26 (0.17; 0.35) \\
      Male & 0.04 (-0.01; 0.09) & 0.21 (0.13; 0.30) & 0.28 (0.23; 0.33)\\
      Education low (high) & 0.08 (-0.25; 0.40) & -0.43 (-0.54; -0.32) & -1.91 (-2.18: -1.64) \\ 
      Education medium (high) & 0.28 (0.06; 0.49) & 0.28 (0.10; 0.45) & -0.62 (-0.74; -0.49)\\
      With partner & 0.16 (-0.01; 0.33) & 0.27 (0.19; 0.36) & 0.36 (0.24; 0.47) \\
      Partner education low (high) & 0.28 (0.09; 0.46) & 0.28 (0.22; 0.33) & 0.28 (0.22; 0.33) \\ 
      Partner education medium (high) & 0.27 (0.04; 0.52) &  0.28 (0.17; 0.39) & 0.28 (0.14; 0.41)\\
      Income low (high) & 0.28 (0.08; 0.47) & -0.10 (-0.16; -0.05) &  -0.76 (-0.87; -0.65)\\
      Income medium (high) & 0.28 (0.23; 0.33) & 0.28 (0.14; 0.41) & 0.07 (0.01; 0.12)\\
      Children under 12 (over 12) & 2.37 (2.29; 2.44) & 1.16 (0.83; 1.49) & 1.48 (1.22; 1.75) \\
      No children (over 12) & 1.74 (1.68; 1.80) & 0.75 (0.50; 1.00) & 1.04 (0.88; 1.20)\\
      Retired & 0.28 (0.03; 0.53) & -0.39 (-0.44; -0.34) & -0.60 (-0.80; -0.40)\\
      Unemployed & 0.67 (0.61; 0.72) &  0.28 (0.18; 0.37) & 0.28 (0.13; 0.42)\\     
      \noalign{\smallskip}\hline\noalign{\smallskip}
    \end{tabular}
    \label{tab:beta}
\end{table}\\Furthermore, Table 9 shows the probability of belonging to the most and to the least digitalized groups for some covariates of interest. It is interesting to notice that the model enables these probabilities to be calculated when varying the country group considered. For example, the probability of belonging to the most digitised group is 0.16 for individuals aged between 65 and 75 in the least digitalized countries, but this probability increases to 0.24 for the most digitalized countries. A similar consideration applies to individuals over 75 years of age. Furthermore, the probability of belonging to the digitalized group is 0.40 for healthy individuals in the least digitised countries, and rises to 0.52 for the most digitalized countries. Furthermore, we can notice that individuals without children are about 20\% more likely to belong to the most digitalized group in both country groups.
\begin{table}[ht!]
\footnotesize
    \centering
     \caption{Probability of belonging to group $g$=1, 2 for some covariates of interest.}
    \tiny{
    \begin{tabular}{{lcccccccccccccc}}
    \toprule
         &  \multicolumn{14}{c}{Variables}\\
        &  \multicolumn{3}{c}{Age} &  \multicolumn{3}{c}{Health}   &  \multicolumn{3}{c}{Education}   &  \multicolumn{3}{c}{Income}   &  \multicolumn{2}{c}{Children}\\
        \cmidrule(lr){2-4}
        \cmidrule(lr){5-7}
        \cmidrule(lr){8-10}
        \cmidrule(lr){11-13}
        \cmidrule(lr){14-15}
       $g$=1  & $<65$ & $65\vdash75$ & $75+$ & Bad & Fair & Good & Low & Medium & High & Low & Medium & High & Yes & No \\
       \midrule
       $q$=1 & 0.24 & 0.84 & 0.92 & 0.63 & 0.77 & 0.60 & 0.97 & 0.91 & 0.12 & 0.92 & 0.84 & 0.24 & 0.21 & 0.79  \\
       $q$=2 & 0.38 & 0.76 & 0.86 & 0.87 & 0.65 & 0.48 & 0.94 & 0.86 & 0.20 & 0.86 & 0.75 & 0.39 & 0.25 & 0.75 \\
      \midrule
      $g$=4  & $<65$ & $65\vdash75$ & $75+$ & Bad & Fair & Good & Low & Medium & High & Low & Medium & High & Yes & No \\
       \midrule
       $q$=1 & 0.76 & 0.16 & 0.08 & 0.37 & 0.23 & 0.40 & 0.03 & 0.09 & 0.88 & 0.08 & 0.16 & 0.76 & 0.79 & 0.21\\       
       $q$=2 & 0.62 & 0.24 & 0.14 & 0.13 & 0.35 & 0.52 & 0.06 & 0.14 & 0.80 & 0.14 & 0.25 & 0.61 & 0.75 & 0.25\\
         \bottomrule
    \end{tabular}}
    \label{etaa}
\end{table}


\section{Conclusions}\label{sec:conclusions}
The mixture of latent trait analyzers (MLTA) is modified to achieve a threefold objective for the analysis of bipartite networks: i) performing a clustering of sending nodes; ii) explicitly modelling the effect that concomitant variables may
have on the clustering structure; iii) properly account for the hierarchical structure of the data; iv) and obtain a clustering of layers. Furthermore, the model originally allows the dependence between receiving nodes to be modelled via a multi-dimensional continuous latent trait. The simulation study shows that the model can be effectively employed
for clustering bipartite networks. In detail, when the number of sending and receiving nodes increases, the proposal is able to correctly identify the model parameters and the classification is good. The model is also applied to the European Social Survey (ESS) data set, in order to identify clusters of individuals with a similar level of digitalization, also taking into account the multilevel structure of the data (i.e., individuals nested within countries) and the influence of nodal attributes on the individual level of digitalization. Results show that the digitalization level is strongly influenced by age, income and education level, as was already evident, as well as by the presence of children in the household. Furthermore, the use of a non-parametric maximum likelihood approach on the random effect allows countries to be classified in terms of the baseline attitude to digital technologies of their residents. It is worth noting that Internet use is a pre-requisite for some of the other skills. Furthermore, a group that does not use technology at all does not emerge. This may be due to the fact that the use of the Internet is too little "distant" from the group with low skills. Both the simulation study and the real data analysis have been carried out via the open-source R software (R Core Team , 2021). All the codes and the data used in this paper are available upon request.\\From a methodological point of view, the proposed approach can be extended to properly handle response variables with more than two categories, as well as in context with more than two levels of clustering, concomitant variables at each level and varying effects of variables at lower levels within groups of layers.

\section*{Statements and Declarations}
A version of the manuscript is in production at Multivariate Behavioral Research.


\begin{thebibliography}{}

\bibitem[Abendroth et al.(2023)]{bollettino} Abendroth, A., Lükemann, L., Hargittai, E., Billari, F., Treas, J., \& van der Lippe, T. (2023). \emph{Digital Social Contacts in Work and
Family Life: Topline results from Round 10 of the European Social Survey}.

\bibitem[Agresti(2002)]{agresti} Agresti, A. (2002). \emph{Categorical Data Analysis}. New York: Wiley.

\bibitem[Alves et al.(2019)]{alves} Alves, L.G.A., Mangioni, G., Cingolani, I. et al. (2019). The nested structural organization of the worldwide trade multi-layer network. \emph{Scientific Reports}, \emph{9}, 2866.

\bibitem[Astegiano et al.(2017)]{astegiano} Astegiano, J., Altermatt, F. \& Massol, F. (2017). Disentangling the co-structure of multilayer interaction networks: degree distribution and module composition in two-layer bipartite networks. \emph{Scientific Reports}, \emph{7}, 15465.

\bibitem[Bartholomew et al.(2011)]{bart} Bartholomew, D.~J., Knott, M., \& Moustaki, I. (2011). \emph{Latent variable models and factor analysis: A unified approach}. Chichester, West Sussex: Wiley.

\bibitem[Davis et al.(2003)]{davis2} Davis, G. F., Yoo, M., \& Baker, W. E. (2003). The Small World of the American Corporate Elite, 1982-2001. \emph{Strategic Organization, 1}(3), 301--326. 


\bibitem[Dempster et al.(1977)]{dempster} Dempster, A.~P., Laird, N.~M., \& Rubin, D.~B. (1977). Maximum Likelihood from Incomplete Data via the EM Algorithm. \emph{Journal of the Royal Statistical Society. Series B (Methodological)}, \emph{39}(1), 1–-38.

\bibitem[Efron(1979)]{efronboot} Efron, B. (1979). Bootstrap Methods: Another Look at the Jackknife. \emph{The Annals of Statistics} \emph{7}(1), 1--26.

\bibitem[ESS round 10 (2020)]{ess10} ESS Round 10: European Social Survey Round 10 Data (2020). Data file edition 3.1. Sikt  -Norwegian Agency for Shared Services in Education and Research, Norway - Data Archive and distributor of ESS data for ESS ERIC. doi:10.21338/NSD-ESS10-2020.

\bibitem[European Commission(2022)]{desi} European Commission. (2022). Digital Economy and Society Index (DESI) reports.

\bibitem[Failli et al.(2022)]{failli} Failli, D., Marino, M.F., \& Martella, F. (2022). \emph{Extending finite mixtures of latent trait analyzers for bipartite networks}. In Balzanella A., Bini M., Cavicchia C. and Verde R. (Eds.) Book of short Paper SIS 2022 (pp. 540-550), Pearson.

\bibitem[Goldstein(2011)]{goldstain} Goldstein, H. (2011). Bootstrapping in multilevel models. In J. J. Hox \& J. K. Roberts (Eds.), Handbook for advanced multilevel analysis (pp. 163–171). Routledge/Taylor \& Francis Group.

\bibitem[Gollini \& Murphy(2014)]{gol2014} Gollini, I., \& Murphy, T.~B. (2014). Mixture of latent trait analyzers for model-based clustering of categorical data. \emph{Statistics and Computing}, \emph{24}(4), 569--588.

\bibitem[Gollini(2020)]{gol2020} Gollini, I. (2020). A mixture model approach for clustering bipartite networks. \emph{Challenges in Social Network Research Volume in the Lecture Notes in Social Networks (LNSN - Series of Springer)}


\bibitem[Hidalgo \& Hausmann(2009)]{hidalgo} Hidalgo, C. \& Hausmann, R. (2009). The building blocks of economic complexity \emph{Proc. Nat. Acad. Sci. 26}, 10570--10575


\bibitem[Jaakkola \& Jordan(1997)]{jaakkola} Jaakkola, T.~S., \& Jordan, M. I. (1997). Bayesian logistic regression: A variational approach. In Madigan, D., Smyth, P. (Eds.) \emph{Proceedings of the 1997 Conference on Artificial Intelligence and Statistics}. Ft. Lauderdale, FL.

\bibitem[Koptelov et al.(2021)]{koptelov} Koptelov, M., Zimmermann, A., Crémilleux, B., \& Soualmia, L. (2021). LPbyCD: a new scalable and interpretable approach for Link Prediction via Community Detection in bipartite networks. \emph{Applied Network Science}, \emph{6}, 10.1007/s41109-021-00415-1. 

\bibitem[Laird(1978)]{npml} Laird, N. (1978). \emph{Nonparametric maximum likelihood estimation of a mixing distribution}. Journal of the Amrican Statistical Association, 73, 805-811.

\bibitem[Pavlopoulos et al.(2018)]{bio} Pavlopoulos, G.A., Kontou, P.I., Bouyioukos, C., Markou, E., Bagos, P.G. (2018). Bipartite graphs in systems biology and medicine: a survey of methods and applications. \emph{Gigascience}, \emph{7}(4), 1--31.


\bibitem[Rand(1971)]{rand} Rand, W.M. (1971). Objective Criteria for the Evaluation of Clustering Methods. \emph{Journal of the American Statistical Association}, \emph{66}, 846--850.

\bibitem[Schwarz(1978)]{schwarz} Schwarz, G. (1978). Estimating the Dimension of a Model. \emph{Annals of Statistics}, \emph{6}, 461--464.

\bibitem[Vermunt \& Magidson(2005)]{vermunt}Vermunt, J. K., \& Magidson, J. (2005). Hierarchical mixture models for nested data structures. In C. Weihs, \& W. Gaul (Eds.), Classification: The Ubiquitous Challenge, 176--183. Springer.

\end{thebibliography}
\end{document}